\RequirePackage{lineno}

\documentclass[aps, 10.5pt, superscriptaddress, twocolumn, nofootinbib]{revtex4-2}
\usepackage[UKenglish]{babel}
\usepackage{latexsym}
\usepackage[pdftex, colorlinks=true, linkcolor=red, citecolor=blue, urlcolor=magenta, pdfstartview=FitH]{hyperref}
\usepackage{graphicx}	
\usepackage{amssymb}
\usepackage{amsmath}
\usepackage{xcolor}
\usepackage[normalem]{ulem}

\pdfoutput=1

\usepackage{bm}
\usepackage{verbatim}
\usepackage{amsmath}
\usepackage{amssymb}
\usepackage[utf8]{inputenc}
\usepackage{nicefrac}
\usepackage{braket}
\usepackage{pdfcomment}

\usepackage{pbox}
\usepackage{makecell}
\usepackage{hhline}
\usepackage{multirow}
\usepackage{tabularx}
\usepackage{soul}
\DeclareUnicodeCharacter{0308}{HERE!HERE!}

\DeclareUnicodeCharacter{2212}{\textendash}

\begin{document}
\title{Strain control of exciton and trion spin-valley dynamics in monolayer transition metal dichalcogenides}
%
%
\author{Z. An}
\thanks{Contributed equally to this work.}
\affiliation{Institute of Solid State Physics, Leibniz Universit\"at Hannover, Appelstraße 2, 30167 Hannover, Germany.}
\author{P. Soubelet}\email{pedro.soubelet@wsi.tum.de}
\thanks{Contributed equally to this work.}
\affiliation{Walter Schottky Institut and TUM School of Natural Sciences, Technische Universit\"at M\"unchen, Am Coulombwall 4, 85748 Garching, Germany.}
\author{Y. Zhumagulov}
\affiliation{Institute for Theoretical Physics, University of Regensburg, 93040 Regensburg, Germany.}
\author{M. Zopf}\email{michael.zopf@fkp.uni-hannover.de}
\affiliation{Institute of Solid State Physics, Leibniz Universit\"at Hannover, Appelstraße 2, 30167 Hannover, Germany.}
\author{A. Delhomme}
\affiliation{Walter Schottky Institut and TUM School of Natural Sciences, Technische Universit\"at M\"unchen, Am Coulombwall 4, 85748 Garching, Germany.}
\author{C. Qian}
\affiliation{Walter Schottky Institut and TUM School of Natural Sciences, Technische Universit\"at M\"unchen, Am Coulombwall 4, 85748 Garching, Germany.}
\author{P. E. Faria~Junior}
\affiliation{Institute for Theoretical Physics, University of Regensburg, 93040 Regensburg, Germany.}
\author{J. Fabian}
\affiliation{Institute for Theoretical Physics, University of Regensburg, 93040 Regensburg, Germany.}
\author{X. Cao}
\affiliation{Institute of Solid State Physics, Leibniz Universit\"at Hannover, Appelstraße 2, 30167 Hannover, Germany.}
\author{J. Yang}
\affiliation{Institute of Solid State Physics, Leibniz Universit\"at Hannover, Appelstraße 2, 30167 Hannover, Germany.}
\author{A. V. Stier}
\affiliation{Walter Schottky Institut and TUM School of Natural Sciences, Technische Universit\"at M\"unchen, Am Coulombwall 4, 85748 Garching, Germany.}
\author{F. Ding}
\affiliation{Institute of Solid State Physics, Leibniz Universit\"at Hannover, Appelstraße 2, 30167 Hannover, Germany.}
\author{J. J. Finley}
\affiliation{Walter Schottky Institut and TUM School of Natural Sciences, Technische Universit\"at M\"unchen, Am Coulombwall 4, 85748 Garching, Germany.}
%
%
\date{\today}
%
%

\begin{abstract}
The electron-hole exchange interaction is a fundamental mechanism that drives valley depolarization via intervalley exciton hopping in semiconductor multi-valley systems. Here, we report polarization-resolved photoluminescence spectroscopy of neutral excitons and negatively charged trions in monolayer MoSe$_2$ and WSe$_2$ under biaxial strain. We observe a marked enhancement(reduction) on the WSe$_2$ triplet trion valley polarization with compressive(tensile) strain while the trion in MoSe$_2$ is unaffected. The origin of this effect is shown to be a strain dependent tuning of the electron-hole exchange interaction. A combined analysis of the strain dependent polarization degree using \textit{ab initio} calculations and rate equations shows that strain affects intervalley scattering beyond what is expected from strain dependent bandgap modulations. The results evidence how strain can be used to tune valley physics in energetically degenerate multi-valley systems.
\end{abstract}

%
%
\maketitle
%
%


Semiconducting transition metal dichalcogenides (TMDs) are layered materials with strong light-matter interactions. In the monolayer (ML) limit, they are direct bandgap materials~\cite{Mak.2010, Splendiani2010} at the $K/K'$ points of their hexagonal Brillouin zone~\cite{Xiao.2012}, where interband optical transitions form tightly bound excitons~\cite{Mak2013, chernikov2014excitons, Stier.2016, Stier.2018, Goryca.2019}. Furthermore, strong spin-orbit coupling (SOC) and the inherently broken inversion symmetry couples spin- and valley degrees of freedom causing chiral optical selection rules with marked valley dichroism~ \cite{Gunawan2006, xu2014spin, mak2018light, schaibley2016valleytronics, xu2014spin, sallen2012robust, cao2012valley, mak2012control, zeng2012valley}. Based on initial first-principles calculations that predicted long valley coherence times~\cite{PhysRevB.88.085433, zhu2014exciton} and the possibility to coherently control this pseudospin~\cite{wang2016control, Wang2018, schmidt2016magnetic, cao2012valley, mak2012control, hao2016direct, hao2017trion}, the concept of valleytronics was envisioned as a promising route to process and store information. However, short exciton lifetimes, fast decoherence and fast valley depolarization limit practical applications of 2D TMDs for valleytronics~\cite{schaibley2016valleytronics,robert2016exciton, moody2016exciton, zhu2014exciton, yu2014valley}. While different approaches have been pursued to investigate valley depolarization in TMDs, significant variations in the valley depolarization times from a few picoseconds (ps)~\cite{zhu2014exciton, schmidt2016ultrafast, PhysRevB.95.075428, PhysRevB.92.235425} to several tens of ps~\cite{yan2015valley, PhysRevB.90.161302, PhysRevLett.112.047401, plechinger2016trion} have been observed, highlighting the continued need for deeper understanding of the underlying valley physics. 
Of particular interest are the various electron-electron and electron-hole interaction channels in optically bright and dark ML TMDs~\cite{Kormnyos.2015, PhysRevB.106.085414, PhysRevB.101.115307}.   

At cryogenic temperatures, valley depolarization in semiconductor multi-valley systems is mainly driven by electron-hole exchange interaction (EHEI)~\cite{ye2017optical, hao2016direct, yan2015valley, wang2016control, pattanayak2022steady, PhysRevB.89.205303, PhysRevB.47.15776}. This mechanism, depicted schematically by its Feynman diagram for TMDs in figure \ref{figura:1}a, describes the Coulomb interaction between an electron in the conduction band (CB) at $K$, with an electron in the valence band (VB) at $K'$. As a result, the electron in the CB at $K$ is scattered to the VB at $K$ while the electron in the VB of $K'$ is scattered to the CB at $K'$~\cite{PhysRevB.89.205303}. Effectively, the EHEI results in the annihilation of a bright exciton at $K$ and the creation of a bright exciton at $K'$. The effect of the EHEI markedly depends on the TMD bandstructure and exciton configurations~\cite{Kormnyos.2015, Wang2018}. For instance, in ML molybdenum-diselenide (MoSe$_2$), adding an electron to the neutral exciton limits the EHEI and protects trions from intervalley scattering, reducing trion valley depolarization~\cite{yu2014dirac, schaibley2016valleytronics}. In contrast, negatively charged trions in tungsten-diselenide (WSe$_2$) are split into the \textit{inter}valley triplet trion ($T_t$) and the \textit{intra}valley singlet trion ($T_s$) configurations~\cite{Courtade2017,zipfel2020light}, allowing electron-hole pair hopping from $K$ to $K'$ through triplet-to-singlet conversion~\cite{Zhumagulov2022}.

In this Letter, we use piezoelectric devices to apply biaxial strain ($s$) to ML WSe$_2$ and MoSe$_2$ at cryogenic temperatures~\cite{martin2017strain, iff2019strain, ziss2017comparison, ding2010stretchable} and investigate the valley depolarization of excitons and negatively charged trions. Although strain engineering was broadly used in 2D materials to study modulations of the bandgap and vibrational modes~\cite{ding2010stretchable, PhysRevB.106.125303, dadgar2018strain, wang2019situ, Guan2015, Wang2020, Aslan2018, RN83,martin2017strain, iff2019strain, Hui2013, ziss2017comparison, edelberg2020tunable, Cenker2022, hernandez2022strain}, the effect of strain on the valley depolarization was studied theoretically~\cite{aas2018strain, wang2020strain} and the only experimental realization was performed trough uniaxial strain at room temperature~\cite{PhysRevB.88.121301}. By performing circularly polarized photoluminescence (PL) spectroscopy as a function of $s$, we observe that while negative trions in MoSe$_2$ are not affected, triplet trions in WSe$_2$ strongly valley depolarize. \textit{Ab initio} calculations show that $s$ predominantly affects the EHEI and therefore the intervalley scattering time via the modulation of the bandgap. Using these results, we model the strain-dependent exciton/trion polarization with rate equations that consider the interplay between the exciton/trion radiative lifetime and the intervalley scattering time. Although we find qualitative agreement between experiments and theory, the experimentally observed variation of the exciton/trion depolarization is notably stronger. Our observations, therefore, suggest the use of $s$ as an efficient way to control exciton and trion valley dynamics while maintaining $K/K'$-degeneracy. 

\section{results and discussion}
\label{results}

\begin{figure}[t!!]
\includegraphics*[keepaspectratio=true, clip=true, angle=0, width=1.\columnwidth, trim={0mm, 2mm, 173mm, 0mm}]{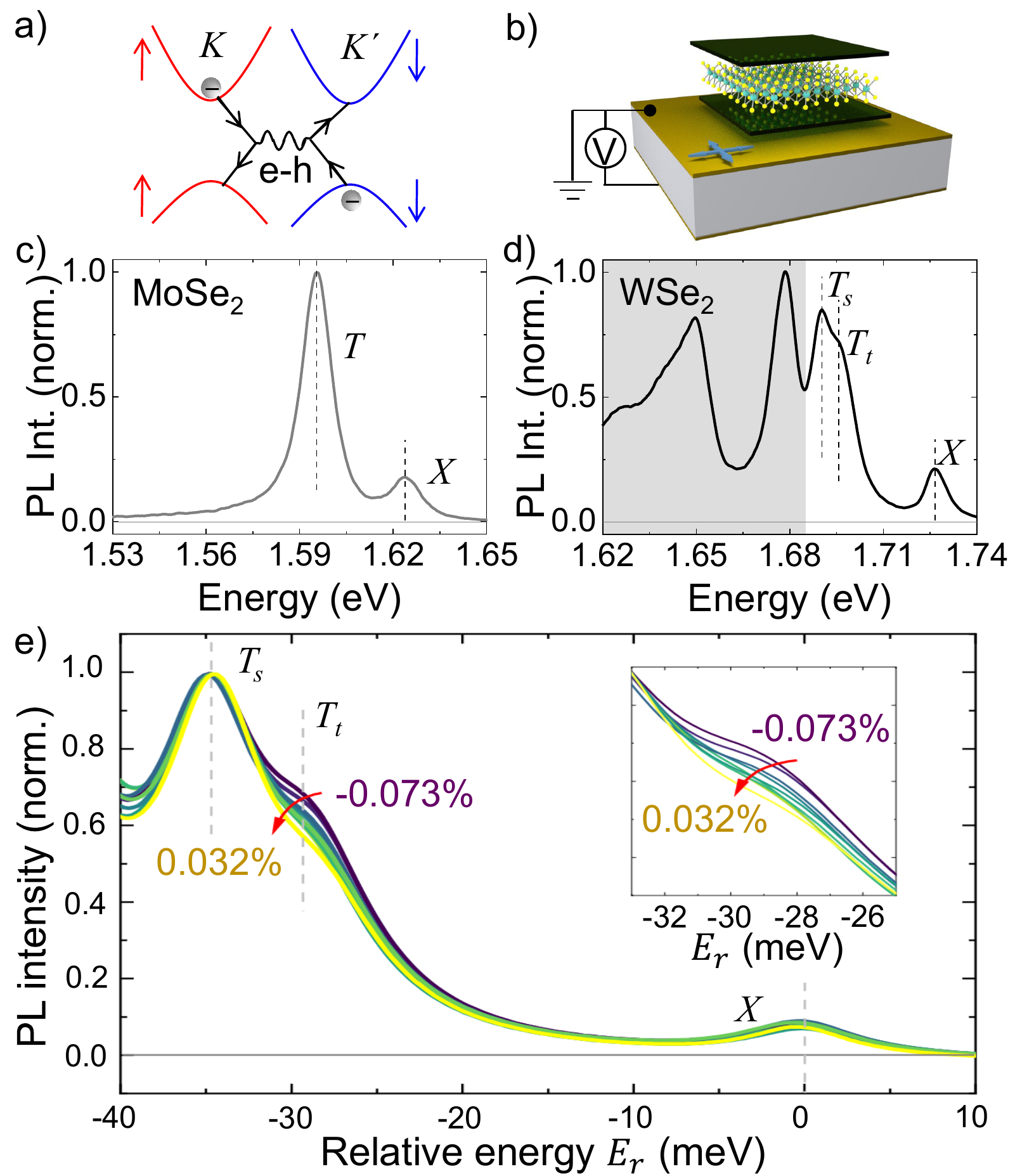}
\caption{\textbf{a)} Feynman diagram of the bright exciton transition between $K$ and $K'$ due to EHEI. A spin up electron in the CB at $K$ and a spin down electron in the VB at $K'$ are scattered to their final states in the VB at $K$ and the CB at $K'$, respectively. \textbf{b)} Schematic of the sample stack: piezoelectric substrate, its electric connections and the hBN encapsulated ML TMD. $\mu$PL spectra at $10\,$K of the \textbf{c)} ML MoSe$_2$ and \textbf{d)} WSe$_2$. The MoSe$_2$ spectra show the $X$ and $T$ emission. In WSe$_2$, the spectra display the $X$, the triplet ($T_t$) and the singlet ($T_s$) trion. Emissions at lower energy (grey shade) are outside the scope of this letter. \textbf{e)} WSe$_2$ PL for $s$ ranging from -0.073\,\% to 0.032\,\%. Energy scale is relative to $X$ and spectra are normalized to the $T_s$ emission intensity. Inset: detail of the PL emission in the $T_t$ spectral region. 
}
\label{figura:1}
\end{figure}

The strain actuators are piezoelectric crystals of lead magnesium niobate-lead titanate (PMN-PT)~\cite{li2019giant} with electrical top/bottom gold contacts. Biaxial strain at cryogenic temperatures ($10$\,K) is applied to the ML TMDs stack by poling the PMN-PT crystal with a constant voltage $V$ across the piezoelectric element, as shown in figure \ref{figura:1}b~\cite{martin2017strain, iff2019strain, ziss2017comparison, ding2010stretchable}. The ML TMDs were obtained from commercial bulk crystals through mechanical exfoliation. The ML TMDs were subsequently encapsulated between thin hexagonal boron nitride (hBN) by using dry transfer techniques based on polycarbonate films, similar to Ref.\onlinecite{castellanos2014deterministic}. The assembled structure was stamped directly on top of the piezoelectric substrates. Special care was taken to ground the top electrode of the piezoelectric element to prevent unintentional charging of the TMDs during the experiments. Details about the strain devices are presented in the Supplemental Material (SM) \ref{piezos}.  

Photoluminescence spectra were recorded at the center of each ML TMD with an optical CW power $P = 1~\mu$W focused to a diffraction limited spot (100$\times$ objective, NA = 0.7). The samples were excited near resonance at $E_P = 1.96$\,eV for WSe$_2$ and $E_P = 1.68$\,eV for MoSe$_2$.
All data shown in this Letter were found to be strain reversible (see SM \ref{piezos}). Representative PL spectra for the samples at zero applied strain are shown in Fig.~\ref{figura:1}c and d. The emission feature at $\sim$1.623\,eV($\sim$1.728\,eV) for MoSe$_2$(WSe$_2$), is associated with the neutral bright exciton ($X$), consistent with previous studies for hBN-encapsulated ML TMDs~\cite{cadiz2017excitonic, wierzbowski2017direct, ross2013electrical, PhysRevB.88.045318}. The MoSe$_2$ spectrum shows an additional single peak $\sim 30\,$meV red detuned from the $X$ emission and consistent with negative trions ($T$) of the \textit{inter}valley singlet type~\cite{Wang2018, shepard2017trion, ross2013electrical}. Conversely, WSe$_2$ shows the singlet ($T_s$) and the triplet ($T_t$) trion at $\sim$1.686\,eV and $\sim$1.692\,eV, respectively, in agreement with previous reports~\cite{Wang2018, li2020fine, rivera2021intrinsic, robert2021spin, PhysRevB.105.085302, jones2016excitonic,zipfel2020light}. In addition, we observe the emission from localized states and phonon replicas (grey shade) as described in Ref.\onlinecite{rivera2021intrinsic}. We do not discuss these features further in the remainder of this manuscript, since they are not central to the discussed photophysics. From the relative $T/X$ emission intensity and peak positions, we estimate an electron density $n_{e}($MoSe$_2) = 2.5 \times 10^{10}\,$cm$^{-2}$ and $n_{e}($WSe$_2) = 3 \times 10^{11}\,$cm$^{-2}$ in our samples~\cite{ross2013electrical,robert2021spin}.

We continue by describing the effect of $s$ on the emission energy and intensity of excitons and trions. By varying the voltage applied to the piezoelectric element, from -400\,V to 400\,V, the neutral exciton emission of both materials continuously blue shifts by $\sim$10\,meV, consistent with a total strain variation of $\Delta s \sim 0.1\%$ across the voltage range~\cite{zollner2019strain}. Details about the method used to calibrate $s$ are shown in SM \ref{raw data}. We performed first principles calculations in the range of the experimentally applied strain to confirm that the band gap is mainly affected, whereas changes in the effective masses, spin-mixing, electron-electron interactions~\cite{PhysRevX.13.011029} and interband dipole matrix elements are negligible. For the interested reader, these calculated quantities are summarized in the SM \ref{DFT}.

The effect of strain on the $X/T$ emission intensity is presented in Fig.~\ref{figura:1}e, where we plot the co-polarized PL spectra ($\sigma^+$ excitation and collection) of the WSe$_2$ sample for various strains. In order to highlight the effects of $s$, all spectra are normalized to the singlet trion intensity and the energy axis relative to the $X$ energy ($E_X$)~\cite{dirnberger2021quasi}. $T_s$ and $X$ emission intensities are approximately constant (see SM \ref{raw data} for details), consistent with a picture that exciton generation rate and recombination time are not affected by the small amount of $s$ exerted by the piezoelectric device. However, the triplet trion shows a relative decrease of its intensity by increasing $s$. This experiment shows that the relative intensities between trions is not only sensitive to the electron background density in the sample~\cite{robert2021spin, rivera2021intrinsic, he2020valley} but also to the local strain.

\begin{figure}[t!!]
\includegraphics*[keepaspectratio=true, clip=true, angle=0, width=1.\columnwidth, trim={1mm, 9mm, 173mm, 0mm}]{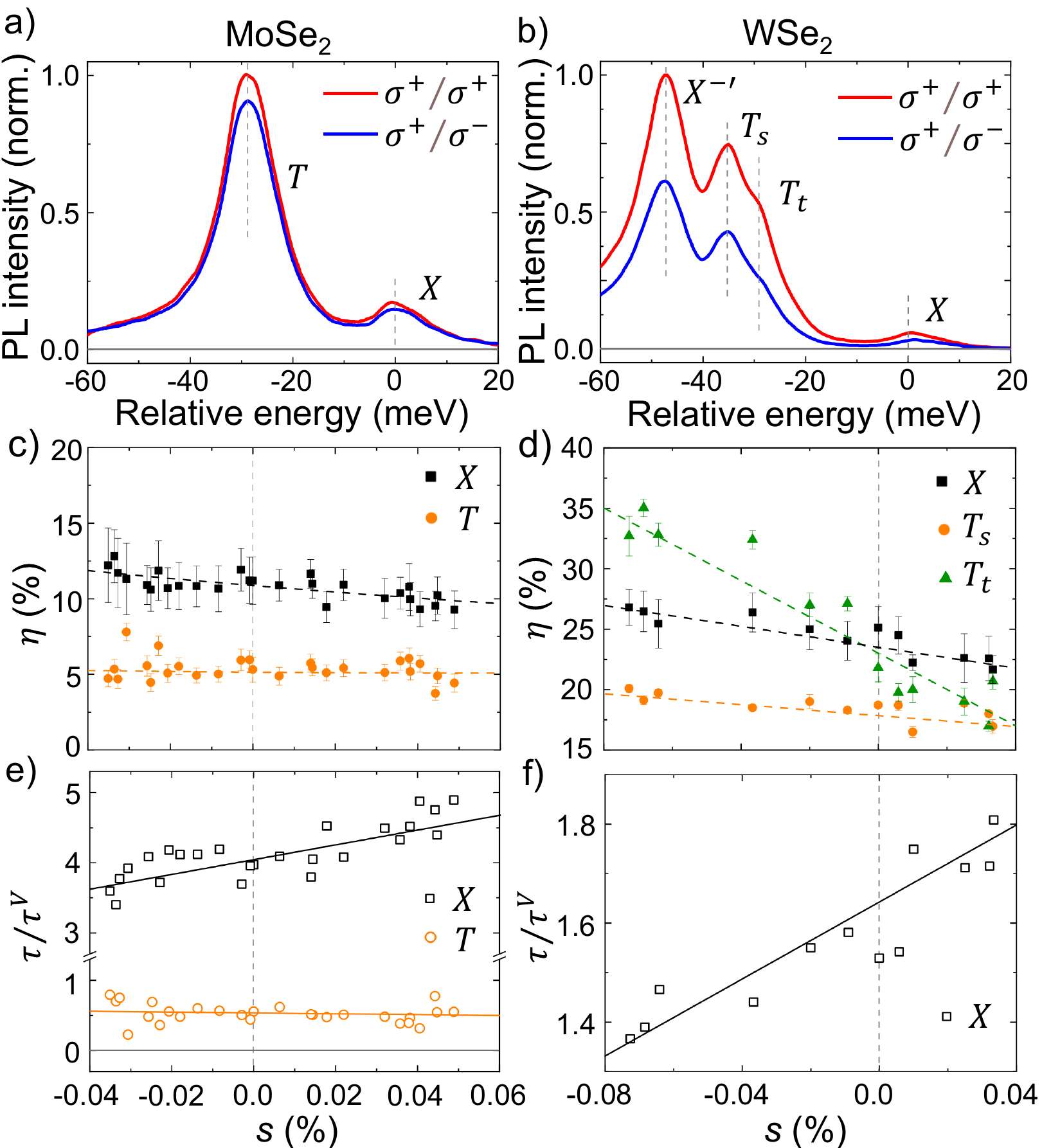}
\caption{Polarization-resolved $\mu$PL spectra for \textbf{a)} ML MoSe$_2$ and \textbf{b)} WSe$_2$ at $s=0$. Red(blue) spectra are obtained for $\sigma^+$ excitation and $\sigma^+$($\sigma^-$) detection. Energy scale is relative to the $X$ and spectra are normalized to the feature with highest intensity in co-polarized configuration. \textbf{c)} $\eta_X$ (black) and $\eta_T$ (orange) for ML MoSe$_2$ as a function of $s$. \textbf{d)} Neutral exciton (black), singlet trion (orange) and triplet trion (green) circular polarization degree as function of $s$ for ML WSe$_2$. Dashed lines in \textbf{c)} and \textbf{d)} are guides to the eye. \textbf{e)} Ratio of total decay rate ($\tau$) to the total intervalley scattering rate ($\tau^V$) for excitons (black dots) and trions (orange dots) in MoSe$_2$. \textbf{f)} Ratio of $\tau/\tau^V$ as a function of $s$ for excitons in WSe$_2$. Solid lines in \textbf{e)} and \textbf{f)} are linear fits to the data.
}
\label{figura:2}
\end{figure}

To further investigate the strain response of $T_t$, we perform polarization-resolved PL as a function of strain on both materials. The spectra at $s=0$ are shown in Fig.~\ref{figura:2}a and b for MoSe$_2$ and WSe$_2$, respectively. Red(blue) correspond to $\sigma^+$ excitation and $\sigma^+$($\sigma^-$) collection. In both cases, the energy scale is relative to $E_X$ and the spectra are normalized to the $\sigma^+$ PL intensity of $T$ for MoSe$_2$ or the peak labeled as $X^{-'}$ for WSe$_2$~\cite{rivera2021intrinsic}. The significant cross-polarized PL intensity suggests a strong valley depolarization mechanism, particularly for MoSe$_2$. We characterize this effect by means of the circular polarization degree ($\eta$) calculated for each feature, $\eta_x = (I^+_x-I^-_x)/(I^+_x+I^-_x)$, where $x$ labels the feature (exciton or trion) and $I^+_x$($I^-_x$) are co-(cross-) polarized intensities obtained by Lorentzian fits to the data. Section \ref{fitting} in the SM provides information about this procedure.  

Figure \ref{figura:2}c shows $\eta_X$ and $\eta_T$ as a function of $s$ for MoSe$_2$ and Fig.~\ref{figura:2}d $\eta_X$ $\eta_{T_s}$ and $\eta_{T_t}$ for WSe$_2$. The difference between these two materials, as well as the different behaviour of excitons and trions are the main experimental observations of this Letter. For MoSe$_2$, $\eta_X$ shows a small but clear variation of $\sim 2\%$ while $\eta_T$ is constant within our experimental error. For WSe$_2$, $\eta_X$ and $\eta_{T_s}$ vary similarly by $\sim 5\%$ while $\eta_{T_t}$ shows a much stronger response, changing by $\sim 15\,\%$ over the whole range of strain investigated. Though these values depend on the excitation power, the general trends are independent of the laser intensity (see the SM \ref{IntWSe2}). The error bars given in Fig.~\ref{figura:2}c and d take the mathematical error of the fitting routine and the polarization accuracy of our setup ($\sim 98\,\%$) into account. 

To understand the strain dependent $\eta$, we model the system by rate equations, considering the strain dependent recombination times ($\tau_{X,T}$) and valley scattering times ($\tau_{X,T}^V$). In the linear regime, the $X$ formation rate in the $K$-valley is proportional to $P$ and $\eta_X$ is then given by (see SM \ref{rate-equations-mose2} and Ref.\onlinecite{mak2012control})
\begin{equation}
\label{eq:1main}
    \eta_X = \frac{\eta_{X0}}{1+{2\tau_X/{\tau^{V}_{X}}}},
\end{equation}
where $\eta_{X0}$ is the spin polarization at the instance of generation. Although $\eta_{X0}$ depends on the \textit{excess energy} $\Delta E = (E_P - E_X)/2$~\cite{aas2018strain, PhysRevLett.121.167401, kioseoglou2016optical}, it is a proportionality factor and does not affect the general tendency we attempt to describe. Therefore, we assume near resonant excitation and $\eta_{X0}=1$. By combining Eq.~\ref{eq:1main} with the $\eta_X$ data (Fig.~\ref{figura:2}c and d), we calculate the ratio $\tau_X/\tau^{V}_{X}$, shown in black squares in figure \ref{figura:2}e and f for MoSe$_2$ and WSe$_2$, respectively. Solid lines are linear fits that yield the strain dependence of $\tau_X/\tau^{V}_{X}$ as $10 \pm 2$\,/\% for MoSe$_2$ and $4 \pm 1$\,/\% for WSe$_2$, i.e. similar relative variation in both cases. 

We continue by describing the strain dependent valley depolarization of trions. For trions in MoSe$_2$, we use the same assumptions as for excitons, which results in (see SM \ref{rate-equations-mose2} and Ref.\onlinecite{mak2012control}) 
\begin{equation}
\label{eq:2main}
    \eta_T = \frac{\eta_{X}}{1+{2\tau_T/{\tau^{V}_{T}}}}.
\end{equation}
Note that as trions require the existence of an exciton, the trion polarization at the instance of generation is $\eta_{X}$. Once again, by combining Eq.~\ref{eq:2main} with the $\eta(s)$ data in figure \ref{figura:2}c we obtain the ratio $\tau_T/\tau^{V}_{T}$ shown as orange circles in figure \ref{figura:2}e. The ratio $\tau_T/\tau^{V}_{T} < 1$ implies comparatively shorter recombination times and dismisses the intervalley scattering influence in the trion dynamics. 

\begin{figure}[t!!]
\includegraphics*[keepaspectratio=true, clip=true, angle=0, width=1.\columnwidth, trim={5mm, 58mm, 205mm, 0mm}]{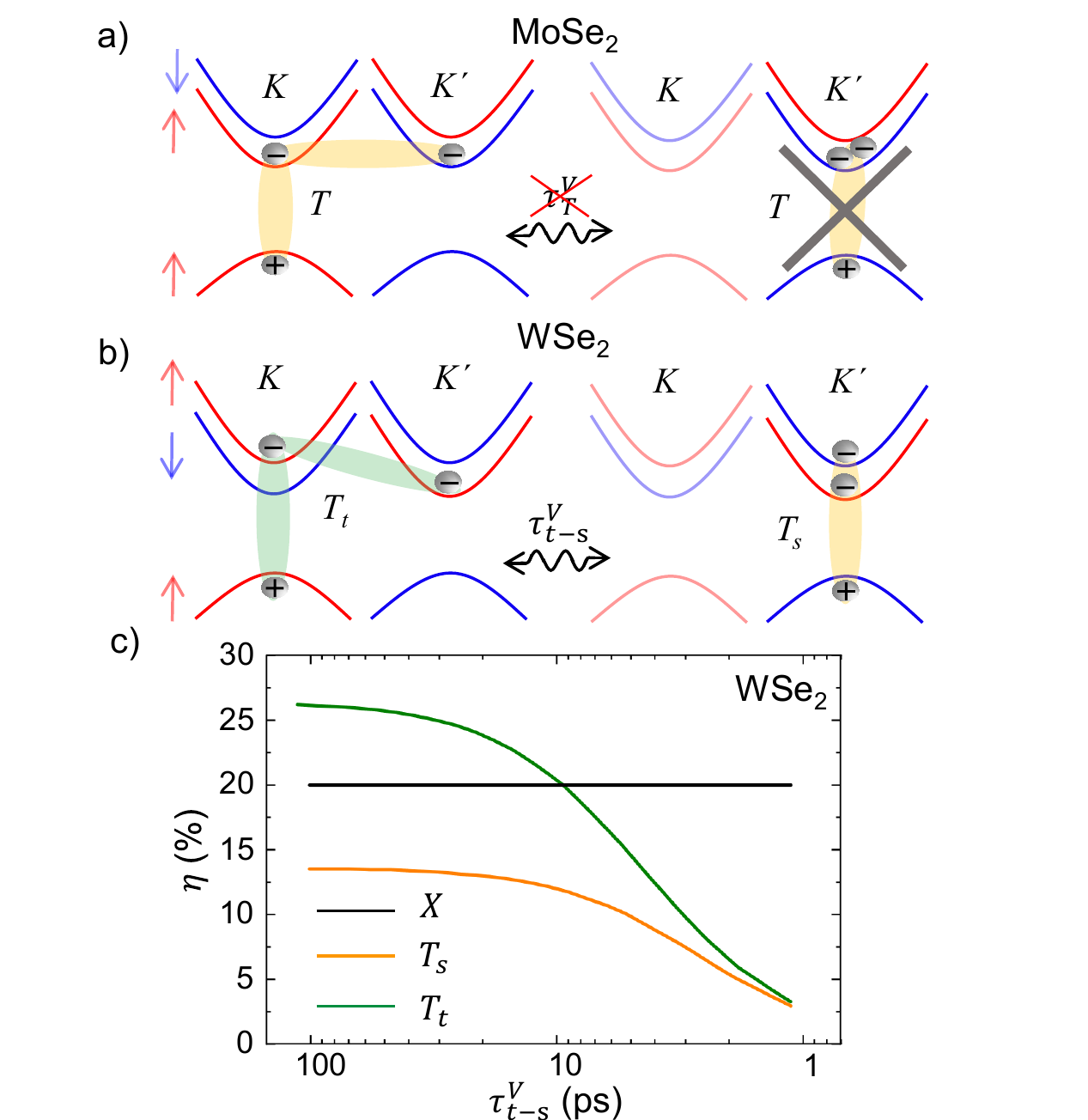}
	\caption{Sketch of the intervalley scattering effects in monolayer TMDs. \textbf{a)} MoSe$_2$ trions are spin protected against EHEI. \textbf{b)} In WSe$_2$, EHEI transforms triplet trions at $K$ into singlet trions at $K'$. \textbf{c)} Calculated $\eta_X$ (black), $\eta_{T_s}$ (orange) and $\eta_{T_t}$ (green) as a function of trion intervalley scattering time $\tau_{t-s}^V$ in WSe$_2$.
 }
\label{figura:3}
\end{figure}

Comparing the ratio $\tau_X/\tau^{V}_{X}$ on both materials, $\tau_X/\tau^{V}_{X}$ is approximately $2.5\times$ larger in MoSe$_2$ than in WSe$_2$. This is consistent with a relatively faster exciton intervalley scattering in MoSe$_2$, which is anticipated due to an additional depolarization mechanisms, such as the Rashba-type mixing of bright and dark excitons~\cite{PhysRevB.101.115307}. In Ref.~\onlinecite{aas2018strain}, the authors calculated $\eta_X$ through a two-band $k\cdot p$ method and found that $\Delta E > 0$ provides a strain dependent valley depolarization. Nevertheless, the strain induced depolarization we observe is two orders of magnitude larger than their prediction, suggesting a different cause. While $\tau_X/\tau^{V}_{X}$ depends on strain, $\tau_T/\tau^{V}_{T}$ is constant across the strain range. We explain this observation by the lack of EHEI for trions in this material since they are spin protected~\cite{hao2017trion, shepard2017trion} (see Fig.~\ref{figura:3}a). Therefore, the strong strain modulation of $\tau_X/\tau^{V}_{X}$ on both materials as well as the lack of strain dependency on $\tau_T/\tau^{V}_{T}$ suggests that $s$ is able to tune the EHEI. For this reason, we model the strain dependent exciton radiative decay rates and intervalley EHEI scattering times from first principles (see SM \ref{exc-val-time} and SM \ref{rad-time}). Our calculations show that the radiative decay time is barely strain dependent. On the other hand, the EHEI strongly depends on the bandgap, which shrinks with strain. Consequently, the EHEI scattering time markedly decreases by increasing strain. Both observations are consistent with our experimental results that, however, display a strain dependence an order of magnitude larger than our calculations. As the small amount of strain applied in the experiments affects the TMDs bandgap but not their bandstructure (see SM \ref{DFT}), any other channel for recombination and intervalley scattering is treated as a constant background without affecting our results.

Finally, we discuss the singlet/triplet depolarization mechanism in WSe$_2$. In Ref.~\onlinecite{robert2021spin}, the authors propose a mechanism based on spin-valley pumping of resident \textit{electrons} for tungsten based TMDs that leads to a pump- and doping-dependent $\eta_{T_{s,t}}$. This effect is based on an efficient phonon mediated scattering of electrons from the upper CB in $K$ to the lower CB in $K'$. For a single pump power, we note that we can interpret our trion results based on this effect by introducing a strain dependent \textit{electron} intervalley scattering time in our rate equations. However, our observations for \textit{excitons} in both materials can only be explained by a strain dependent EHEI. Furthermore, we observe a pump power dependence of $\eta_{T_{s,t}}$ that is different to that described in Ref.~\onlinecite{robert2021spin} (see SM \ref{IntWSe2}). For these reasons, we interpret our strain dependent trion data through a triplet-to singlet conversion mediated by a strain dependent EHEI. The addition of this scattering channel, characterized by a scattering time $\tau_{t-s}^V$, renders the solution of the rate equations non-analytical (see SM \ref{rate-equations-wse2}). As the singlet-to-triplet scattering term is proportional to the singlet and triplet population difference, the resulting populations are interdependent. 

 We numerically solve the rate equations and plot in figure \ref{figura:3}c the resulting values for $\eta_X$, $\eta_{T_s}$ and $\eta_{T_t}$ using a fixed set of realistic input parameters ($\tau_X=1$\,ps, $\tau_{T_t}=2$\,ps, $\tau_{T_s}=4$\,ps, $\tau^V_X =0.5$\,ps and trion formation time $\tau_b=0.1$\,ps) while sweeping $\tau_{t-s}^V$. By decreasing $\tau_{t-s}^V$, the singlet to triplet conversion starts to dominate the intervalley dynamics and $\eta_{T_{s,t}}$ decrease with a markedly different slope ultimately tending to zero. When $\tau_{t-s}^V \sim 10\,$ps our rate equations reproduce the experimentally observed $\eta(s)$ very well (see Fig.~\ref{figura:2}d).

In summary, we observe clear experimental evidence of the EHEI dependence with strain and support our claims with first principle calculations. We presented a detailed study of the biaxial strain impact on the circular polarization degree of excitons and trions in MoSe$_2$ and WSe$_2$. The circular polarization degree depends on the total exciton and trion lifetimes and their depolarization rates~\cite{mak2012control}. Whilst the radiative decay is essentially independent of strain, the strain dependent intervalley scattering is the only consistent way to explain our observations on $\eta(s)$. This suggests that strain modulates the EHEI to a surprisingly large degree beyond expectations from first principles calculations that only take a strain dependent bandgap into account. Our observations may consolidate the variations of optical polarization degrees and intervalley scattering times reported in the literature and point towards a scattering channel besides the occurrence of resident electrons that facilitates valley depolarization, such as mixing of singlet and triplet trion states \cite{PhysRevB.105.L041302}. Our results highlight the need for further understanding of the spin/valley photophysics in TMDs and point out a possible path to enhance valley polarization towards the development of valleytronics in multi-valley materials.

%
%

\section{Acknowledgements}

We gratefully acknowledge the German Science Foundation (DFG) for financial support via SPP-2244 grant (DI 2013/5-1, FI 947/8-1 and FA 971/8-1). P.S., A.D., C.Q., A.V.S. and J.J.F. additionally acknowledge the clusters of excellence MCQST (EXS-2111) and e-conversion (EXS-2089). Z.A., M.Z., X.C., J.Y., and F.D. gratefully acknowledge the European Research Council (No.QD-NOMS GA715770), and the DFG Excellence Strategy–EXC-2123 Quantum Frontiers–39083 7967. Z.A. is funded by the China Scholarship Council. Y.Z., P.E.F.J., and J.F. also acknowledge the financial support of the DFG SFB 1277 (Project-ID 314695032, projects B07 and B11).

Z.A. and P.S. contributed equally to this work. A.V.S., F.D., and J.J.F. conceived the project. P.S. and C.Q. prepared the samples, Z.A. and P.S. performed the optical measurements on WSe$_2$ and P.S. and A.D. on MoSe$_2$. Z.A., M.Z., X.C. and J.Y. provided support on the piezoelectric performance, and M.Z. X.C. and J.Y. supervised the experiments performed in Hannover. Y.Z., P.E.F.J. and J.F. provided support on the theory and \textit{ab initio} calculations. P.S. and Z.A. analysed the data and P.S. and Y.Z. developed the rate equation model. P.S. wrote the paper with input from all co-authors. All authors reviewed the manuscript.

The Authors declare no Competing Financial or Non-Financial Interests.


%
%

\onecolumngrid

\section{Supplemental material}

\subsection{Piezoelectric actuators}\label{piezos}

\begin{figure}[t!!]
\includegraphics*[keepaspectratio=true, clip=true, angle=0, width=1.\columnwidth, trim={0mm, 90mm, 48mm, 0mm}]{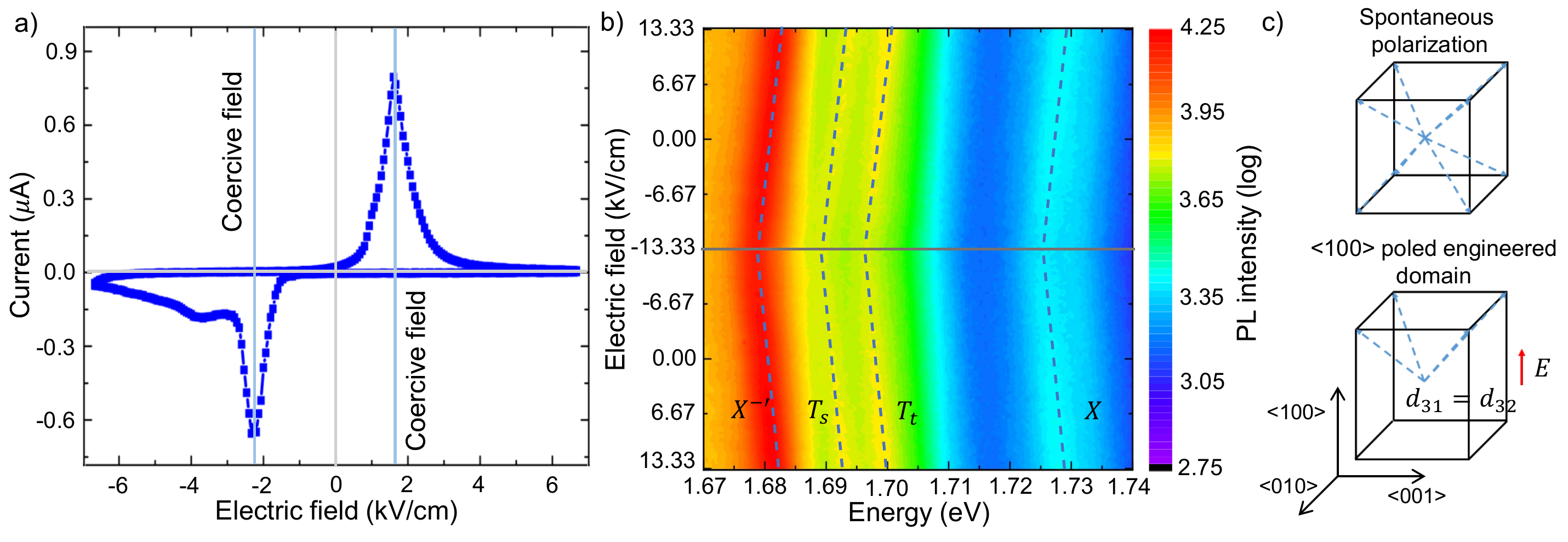}
\caption{Piezoelectric response \textbf{a)} Room temperature characteristic piezoelectric current-voltage response. \textbf{b)} WSe$_2$ PL colour map showing the different excitonic complexes shift as function of electric field onto the piezoelectric actuator. \textbf{c)} PMN-PT crystal in absence (top) and presence (bottom) of an electric field $E$ along the $\langle 100 \rangle$ direction. Dashed lines denote $\langle 111 \rangle$ orientations and the red arrow the electric field. An isotropic in-plane response derives from $d_{31}=d_{32}$ \cite{gao2018giant}.}
\label{figura:SM-1}
\end{figure}

The biaxial piezoelectric substrates consist in a 0.3\,mm thick PMN-PT crystal polished in the $\braket{001}$ crystallographic direction ($z$). The polarization of the crystal along the thickness ($z$) was accomplished by the evaporation of electrodes on both faces of the substrate (20\,nm Cr + 200\,nm Au). The piezoelectric device poling produces a stretching/contraction of the crystal along this direction accompanied by a biaxial contraction/extension of the plane perpendicular to $z$ (biaxial in-plane deformation) \cite{li2019giant, martin2017strain, iff2019strain, ziss2017comparison}.

Figure \ref{figura:SM-1}a shows, at room temperature, the characteristic piezoelectric current-voltage curve. The switching of ferroelectric domains induces the increase of leakage current around $\sim 2\,$kV/cm, defining the coercive field of the actuators at room temperature. The piezoelectric element poling was performed by slowly varying the voltage in order to avoid the switch of ferroelectric domains and resulting creep of strain levels. To perform measurements at low temperatures, the biaxial piezoelectric element was poled at room temperature with 200\,V (6.7\,kV/cm), far above the coercive electric field ($E_c\simeq 1.7$\,kV/cm) and, thereby, the polarization of all the substrate regions was homogeneous. By slowly cooling down the sample, the coercive field of the piezoelectric element increases far above the maximum applied voltage of 500\,V and, as result, all measurements where performed with a unique piezoelectric domain (without switching its polarization) \cite{chen2017temperature}. To perform $\mu$PL experiments as function of strain, the piezoelectric voltage was set to a desired value by slowly sweeping the electric tension. The PL spectra were acquired after some minutes the set voltage was reached in order to allow the mechanical relaxation of the piezoelectric device. The range of $\pm$500\,V correspond to an electric field of $\pm$16.6\,kV/cm applied to the piezoelectric substrate.

Figure \ref{figura:SM-1}b presents, in a colour map, the WSe$_2$ PL measured with $\sim 10\,\mu$W 633\,nm CW laser at 30\,K as function of piezoelectric voltage and emission energy. In the working range, all emission species show linear and reversible shift with the electric field applied onto the piezoelectric actuator.

Strain state consists of two contributions, an arbitrary strain $s_a$ and an active strain $s$, which can be controlled by the piezoelectric actuator. $s_a$ derives from the fabrication process, contraction of the substrate during the cooling down and inhomogeneity or local disorder of the sample. While it differs from point to point, at one position over the piezoelectric element and stable temperature, $s_a$ is constant. We make a coarse estimation of arbitrary strain by using a piezoelectric element on a SiO$_2$/Si wafer. The strain at 30\,K is estimated to be from –0.17\,\% to –0.20\,\%, which means the sample is in a compressive state, in agreement with previous reports \cite{martin2016reversible}. However, the anisotropy of $s_a$ cannot be accessed.

In our research, we focus on the effect of $s$ on monolayer TMDs. The influence of $s_a$ is regarded as an offset. In our case, the piezoelectric single crystal is $\langle 100 \rangle$ cut. During the cooling down, electric tuning is applied along $\langle 100 \rangle$, which presents the engineering domain as presented in figure \ref{figura:SM-1}c.

In the sample fabrication, the TMD flakes are not exactly at the center of the bulk plate, which in principle, brings about asymmetry boundary conditions in the piezoelectric equation. Considering the flake position and high stiffness of PMN-PT, we choose to neglect this minor influence of $s_a$. This modification in the isotropy will not affect the tendency of the optical performance and corresponding interpretation.

\subsection{Details on experiments and data processing}\label{raw data}

\begin{figure}[t!!]
\includegraphics*[keepaspectratio=true, clip=true, angle=0, width=1\columnwidth, trim={-15mm, 62mm, 85mm, 0mm}]{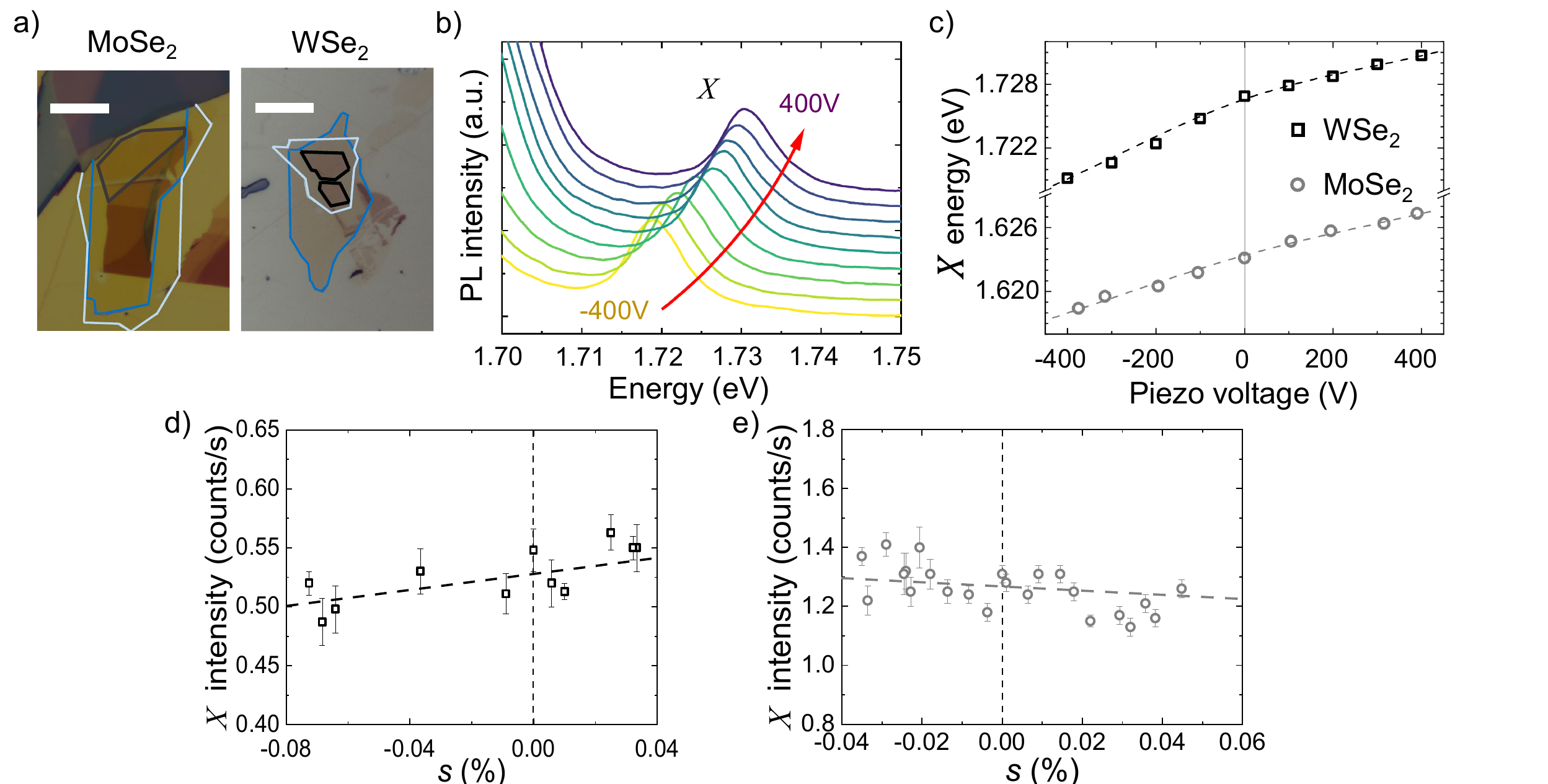}
\caption{\textbf{a)} Optical micrographs of the MoSe$_2$ and WSe$_2$ samples. Blue(light blue) lines indicate the bottom(top) hBN flakes and grey(black) lines the MoSe$_2$(WSe$_2$) monolayer. The white scale bars correspond to 10\,$\mu$m. \textbf{b)} Effect of strain on the WSe$_2$ exciton position for piezoelectric voltages varying between -400\,V to 400\,V. \textbf{c)} WSe$_2$ and MoSe$_2$ exciton position as function of piezoelectric voltage. Dotted lines are guide to the eye. \textbf{d)} and \textbf{e)} Exciton intensity as function of $s$ for WSe$_2$ and MoSe$_2$. Dotted lines are guide to the eye}
\label{figura:SM-7}
\end{figure}

Figure \ref{figura:SM-7}a presents the optical micrograph of the ML WSe$_2$ and MoSe$_2$ samples. Figure \ref{figura:SM-7}b shows shows, as example, the effect of strain on the WSe$_2$ exciton position. By varying the voltage applied to the piezoelectric element from -400\,V to 400\,V, the neutral exciton emission continuously blue shifts by $\sim$10\,meV, consistent with a total strain variation of $\Delta s \sim 0.1\%$ across the voltage range~\cite{zollner2019strain}. We calibrate the strain using the linear relationship $E_X = E_{X(0)} + a_{MX_2} \cdot s$, where $E_{X(0)}$ is the zero strain exciton energy and $a_{MX_2}$ the gauge factor obtained from previous \textit{ab initio} calculations,  namely $a_{MoSe_2} = -98.2\,$meV/\% and $a_{WSe_2} = -133.5\,$meV/\%~\cite{zollner2019strain}. Figure \ref{figura:SM-7}c shows the $X$ fitted emission energy ($E_X$) for WSe$_2$ (black) and MoSe$_2$ (grey) as function of the piezo voltage. Both materials show the expected $X$ blueshift; deviations from strictly linear behaviour are attributed to non-linearities in the piezoelectric response. 

Figure \ref{figura:SM-7}d and e display the exciton intensity for WSe$_2$ and MoSe$_2$, respectively. In both cases, the exciton intensity is approximately constant within the error bars, meaning that the generation efficiency is affected by the detuning of the laser and the exciton energy with strain. 

\subsubsection{Effect of strain on singlet/triplet trion (WSe$_2$)}\label{Raw pol}

\begin{figure}[t!!]
\includegraphics*[keepaspectratio=true, clip=true, angle=0, width=1\columnwidth, trim={-18mm, 65mm, 10mm, 0mm}]{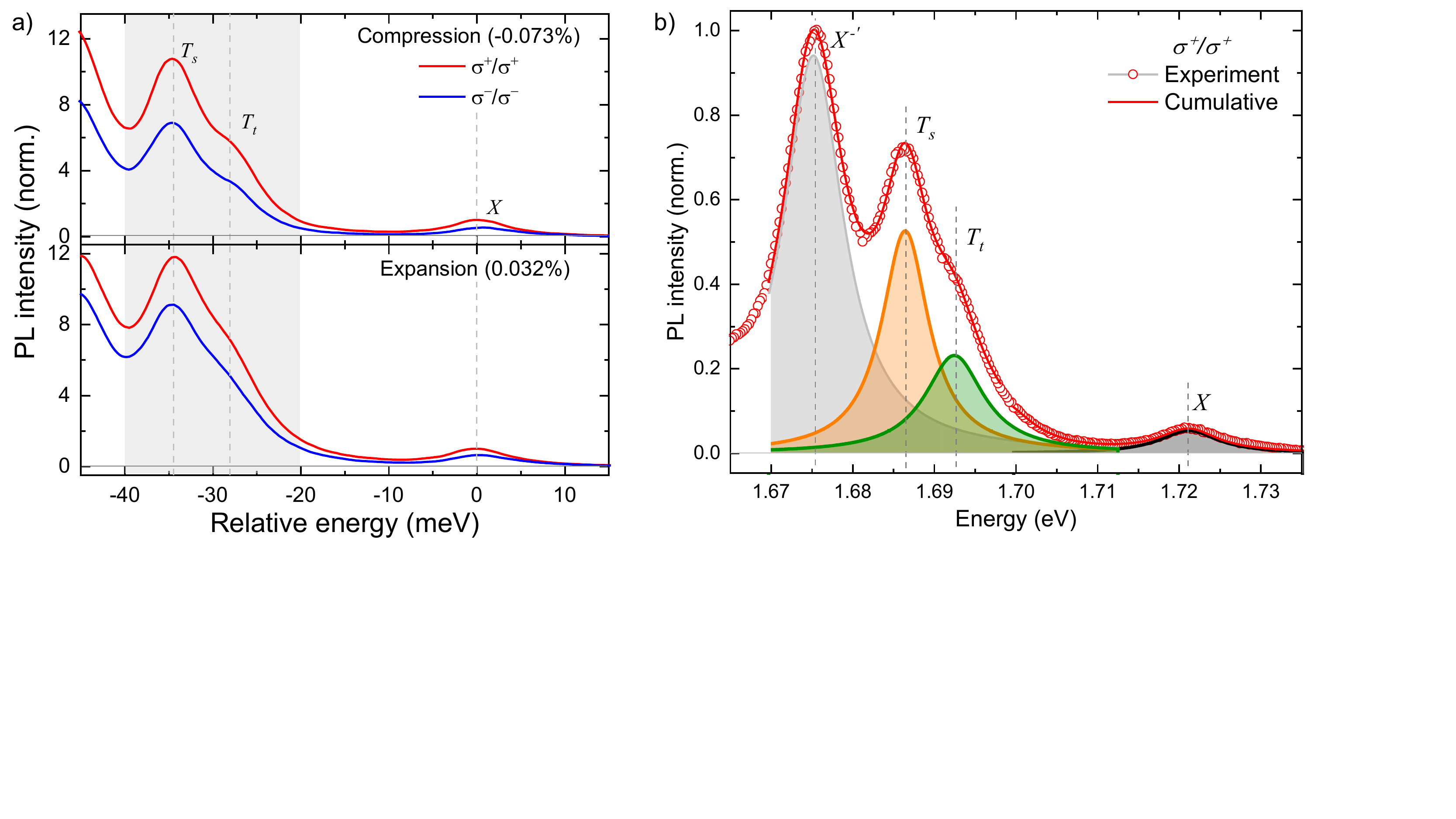}
\caption{\textbf{a)} Co- (red) and cross- (blue) circularly polarized spectra recorded at maximum compression (top panel) and maximum extension (bottom panel) for WSe$_2$. The grey shade highlights the trion region of the spectra where it is possible to observe the main effect of biaxial strain. Energy scale is relative to $E_X$, the spectra are normalized to the co-polarized $X$ peak. \textbf{b)} Lorentzian fits of a typical WSe$_2$ spectrum recorded at zero strain. The figure shows the Lorentzian curves fitted to the peaks labeled as $X^{-'}$(grey), $T_s$(orange), $T_t$(green) and $X$(black). Additionally, the cumulative curve is shown in red.}
\label{figura:SM-2}
\end{figure}

In order to complement the experimental data presented in figure 2 of the main text, figure \ref{figura:SM-2} presents the co- and cross-polarized PL for the maximum negative(positive) biaxial strain on its top(bottom) panel. The light grey shade in the spectral region of $T_s$ and $T_t$ highlights the variation of the emission profile due to $s$. The two extreme cases ($s=-0.073\,\%$ and $s=0.032\,\%$) show, to the naked eye, the changes in the circular polarization degree presented in the main text.

\subsubsection{Fitting procedure}\label{fitting}

The circular polarization degree for excitons and trions presented in the main text was obtained by fitting Lorentzian functions to extract the emission intensity of each excitonic state. 

Figure \ref{figura:SM-2}a presents, as example, the fitting of a typical WSe$_2$ spectra taken at 0V. The figure depicts the good agreement of the Lorentzian curves and the peaks labeled as $X^{-'}$(grey), $T_s$(orange), $T_t$(green) and $X$(black). The fitting was performed in the spectral range from 1.67\,eV to 1.74\,eV to avoid the influence of the emission at lower energy (localized states and phonon replicas \cite{rivera2021intrinsic}).

\subsubsection{Power effect on WSe$_2$}\label{IntWSe2}

\begin{figure}[t!!]
\includegraphics*[keepaspectratio=true, clip=true, angle=0, width=1\columnwidth, trim={1mm, 120mm, 9mm, 0mm}]{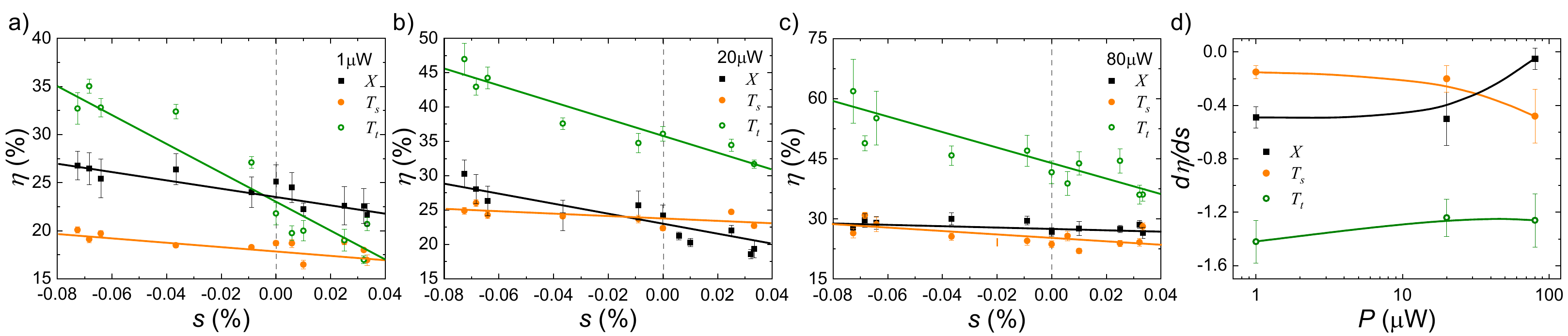}
\caption{WSe$_2$ neutral exciton (black) singlet trion (orange) and triplet trion (green) circular polarization degree as function of strain for different excitation power: \textbf{a)} 1\,$\mu$W, \textbf{b)} 20\,$\mu$W and \textbf{c)} 80\,$\mu$W. \textbf{d)} $d\eta/ds$ as function of pump power. Solid lines in all panels are guide to the eye.}
\label{figura:SM-3}
\end{figure}

Figure \ref{figura:SM-3} presents the circular polarization degree of the analyzed excitonic complexes at different excitation powers. From left to right panel, figure \ref{figura:SM-3} present the experiments performed with an excitation power of 1\,$\mu$W, 20\,$\mu$W and 80\,$\mu$W. In general, the circular polarization degree of each kind of exciton increases with $P$. $\eta_X$ and $\eta_{T_s}$ increases from $\sim25\%$ and $\sim18\%$, respectively, at $P=1\,\mu$W to $\sim25\%$ at $P=20\,\mu$W. However, the total variation of $\eta_x$ as a function of $s$ (slope $d\eta_x/ds$) is approximately constant, as it can be seen in figure \ref{figura:SM-3}d. The experiments performed at higher excitation power (Fig.~\ref{figura:SM-3}c) presents a slightly different tendency for $\eta_X$ and $\eta_{T_s}$, consistent with saturation effects.

\subsection{Rate equation model}

In this section we provide additional information about the rate equation model for TMD monolayers based on Ref.\onlinecite{mak2012control}. In the WSe$_2$ case we add the triplet to singlet trion scattering to describe the trion valley polarization. 

As the bands ordering for MoSe$_2$ and WSe$_2$ is different, we present these cases in two separated subsections.

\subsubsection{MoSe$_2$ case}\label{rate-equations-mose2}

We describe the exciton and trion dynamics in monolayer MoSe$_2$ with six coupled rate equations: two for the density of excitons at $K$ and $K'$, $n_{X}^{K}$ and $n_{X}^{K'}$, respectively, two for the density of trions at $K$ and $K'$, $n_{T}^{K}$ and $n_{T}^{K'}$, respectively, and two extra equations for free electrons in the lower conduction band at $K$ and $K'$, $n_{e}^{K}$ and $n_{e}^{K'}$.

They are:
\begin{align}
\label{rate-mose2}
    \frac{dn_{X}^{K}}{dt}&=P-\frac{n_{X}^{K}}{\tau^r_X}-\frac{n_{X}^{K} n_{e}^{K'}}{C_b}-\frac{n_{X}^{K}-n_{X}^{K'}}{\tau^V_X}\\\nonumber
    \frac{dn_{X}^{K'}}{dt}&=-\frac{n_{X}^{K'}}{\tau^r_X}-\frac{n_{X}^{K'} n_{e}^{K}}{C_b}-\frac{n_{X}^{K'}-n_{X}^{K}}{\tau^V_X}\\\nonumber
    \frac{dn_{T}^{K}}{dt}&=-\frac{n_{T}^{K}}{\tau_T}+\frac{n_{X}^{K} n_{e}^{K'}}{C_b} -\frac{n_{T}^{K}-n_{T}^{K'}}{\tau^V_T}\\\nonumber
    \frac{dN_{T}^{K'}}{dt}&=-\frac{n_{T}^{K'}}{\tau_T}+\frac{n_{X}^{K'} n_{e}^{K}}{C_b} - \frac{n_{T}^{K'}-n_{T}^{K}}{\tau^V_T}\\\nonumber
    \frac{dn_{e}^{K}}{dt}&=\frac{n_{T}^{K'}}{\tau_T}-\frac{n_{X}^{K'} n_{e}^{K}}{C_b}\\\nonumber
    \frac{dn_{e}^{K'}}{dt}&=\frac{n_{T}^{K}}{\tau_T}-\frac{n_{X}^{K} N_{e}^{K'}}{C_b},
\end{align}
where $P$ is the laser pumping of excitons, $\tau^r_X$ the total exciton relaxation time (radiative and non-radiative), $\tau_T$ the trion relaxation time (radiative and non-radiative), $\tau^V_X$($\tau^V_T$) the exciton(trion) valley depolarization time and $C_b$ a trion formation constant such that $C_b/n_{e}^{K(K')} = \tau_b$ the density dependent trion formation time at $K$($K'$). The process of valley-to-valley scattering of excitons is possible due to the pair hopping of an exciton pair from one valley to another. This process conserves the momentum and spin of the exciton. In contrast to excitons, intervalley scattering for low-energy trions is strongly limited due to the single-particle band distribution of trions.

In the steady-state case, the system of equations \eqref{rate-mose2} can be analytically for $(n_{e}^{K}+n_{e}^{K'})/C_b \gg P$ (strong electron doping limit). For excitons, the solution results in circular polarization degree $\eta_{X/T}(t \to \infty)=\left[n_{X}^{K}-n_{X}^{K'}\right]/\left[n_{X}^{K}+n_{X}^{K'}\right]$ as
\begin{align}
    \eta_{X}&=\frac{\tau^{V}_X \left(n_{e} \tau_{X}^r + 2 C_{b}\right)}{n_{e} \tau^{V}_X \tau^r_{X} + 2 C_{b} \tau^{V}_X + 4 C_{b} \tau^r_{X}}.
    \label{eq:mose2-high}
\end{align}
Defining the total exciton decay as $1/\tau_X = 1/\tau^r_X + n_e/C_b$, the solution takes the well known form
\begin{align}
    \eta_{X}=\frac{1}{1 + 2 \tau_{X}/\tau^{V}_X}.
    \label{eq:mose2-highX}
\end{align}
In the case of trions, it results
\begin{align}
    \eta_{T}=\frac{\eta_{X}}{1 + 2 \tau_{T}/\tau^{V}_X}.
    \label{eq:mose2-highT}
\end{align}

\begin{figure}[t!!]
\includegraphics*[keepaspectratio=true, clip=true, angle=0, width=1\columnwidth, trim={50mm, 122mm, 50mm, 0mm}]{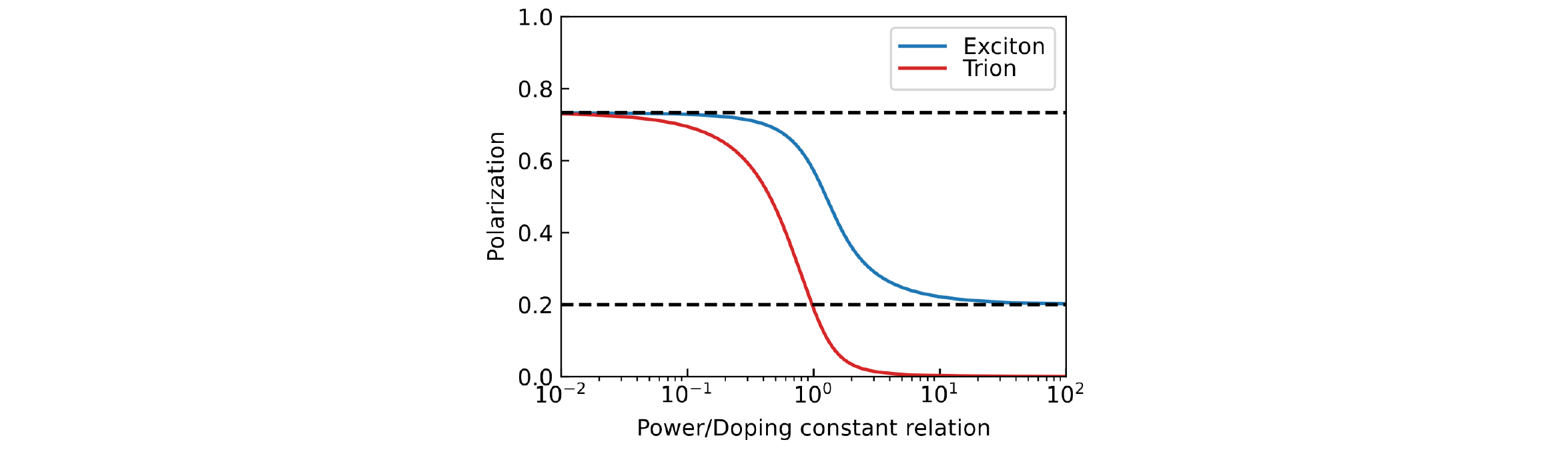}
\caption{Degree of circular polarization for excitons (blue) and trions (red) as function of the ratio of laser pump intensity and electron doping level in a monolayer MoSe$_2$. The following set of parameters were taken as reference realistic values: $\tau^r_X=1$~ps, $\tau_T=2$~ps, $\tau_X^V=0.5$~ps and $C_b=0.1\times 10^{10}$~ps\,cm$^{-2}$. The horizontal dotted lines correspond to the analytical solutions for weak (upper line) and strong (lower line) doping level.}
\label{figura:SM-4}
\end{figure}

In the limit $P\gg(n_{e}^{K}+n_{e}^{K'})/C_b$ (weak electron doping limit), we can obtain, once again, the exciton circular polarization degree $\eta_{X}=1/(1 + 2 \tau_{X}/\tau^V_X)$ and $\eta_{T}=0$. Considering a finite doping level $n_e$ and a laser pumping intensity $P$, the system of equations \eqref{rate-mose2} cannot be analytically solved. The numerical solution is shown in figure \ref{figura:SM-4} and present a decrease in the degree of circular polarization for trions compared to the excitons degree of circular polarization by increasing the ratio of laser pumping $P$ to the electron density $n_e$. Even in the absence of intervalley scattering for trions, the trion degree of circular polarization can be lower than the exciton degree of circular polarization, even more, the presence of trions in the system can increase the exciton circular polarization degree.

\subsubsection{WSe$_2$ case}\label{rate-equations-wse2}

For the monolayer WSe$_2$ case, we used a modified rate equation system compared to the monolayer MoSe$_2$ case. In this material, the monolayer displays two types of trion states: the singlet $n_{T_s}$ and triplet $n_{T_t}$ state. Singlet and triplet trions are formed from a hole and a singlet and triplet electron pair, respectively. In contrast to the MoSe$_2$, the intervalley scattering of trions is possible with the conversion of a triplet trion in a singlet with the opposite circular polarization. This process is possible due to the pair hopping of an exciton pair that form the trion from one valley to the other ($K$ to $K'$ and vice versa), similar to the intervalley exciton scattering process. As for excitons, this process does not violate the momentum and spin conservation law \cite{PhysRevB.89.205303}. Considering the same trion formation time for triplets and singlets, we reflected this differences modifying the system of dynamic equations as follows:
\begin{align}\label{rate-wse2}
    \frac{dn_{X}^{K}}{dt}&=P-\frac{n_{X}^{K}}{\tau^r_X}-n_{X}^{K}\frac{ n_{e}^{K}+n_{e}^{K'}}{C_b}-\frac{n_{X}^{K}-n_{X}^{K'}}{\tau^V_X}\\\nonumber
    \frac{dn_{X}^{K'}}{dt}&=-\frac{n_{X}^{K'}}{\tau^r_X}-n_{X}^{K'}\frac{n_{e}^{K'}+ n_{e}^{K}}{C_b}-\frac{n_{X}^{K'}-n_{X}^{K}}{\tau^V_X}\\\nonumber
    \frac{dn_{T_t}^{K}}{dt}&=-\frac{n_{T_t}^{K}}{\tau_{T_t}}+\frac{n_{X}^{K} n_{e}^{K'}}{C_{b}}-\frac{n_{T_t}^{K}-n_{T_s}^{K'}}{\tau^{V}_{t-s}}\\\nonumber
    \frac{dn_{T_t}^{K'}}{dt}&=-\frac{n_{T_t}^{K'}}{\tau_{T_t}}+\frac{n_{X}^{K'} n_{e}^{K}}{C_b}-\frac{n_{T_t}^{K'}-n_{T_s}^{K}}{\tau^{V}_{t-s}}\\\nonumber
    \frac{dn_{T_s}^{K}}{dt}&=-\frac{n_{T_s}^{K}}{\tau_{T_s}}+\frac{n_{X}^{K} n_{e}^{K}}{C_b}-\frac{n_{T_s}^{K}-n_{T_t}^{K'}}{\tau^{V}_{t-s}}\\\nonumber
    \frac{dn_{T_s}^{K'}}{dt}&=-\frac{n_{T_s}^{K'}}{\tau_{T_s}}+\frac{n_{X}^{K'} n_{e}^{K'}}{C_b}-\frac{n_{T_s}^{K'}-n_{T_t}^K}{\tau^{V}_{t-s}}\\\nonumber
    \frac{dn_{e}^{K}}{dt}&=\frac{n_{T_t}^{K}}{\tau_{T_t}}+\frac{n_{T_s}^{K}}{\tau_{T_s}}-n_{e}^{K}\frac{n_{X}^{K} +n_{X}^{K'}}{C_b}\\\nonumber
    \frac{dn_{e}^{K'}}{dt}&=\frac{n_{T_t}^{K'}}{\tau_{T_t}}+\frac{n_{s}^{K'}}{\tau_{T_s}}-n_{e}^{K'}\frac{n_{X}^{K'} +n_{X}^{K}}{C_b},\\\nonumber
\end{align}
where $n_{T_t}^{K}$ and $n_{T_t}^{K'}$ are the densities of triplet trion states and $n_{T_s}^{K}$ and $n_{T_s}^{K'}$ are the densities of singlet trion states at $K$ and $K'$, respectively. The characteristic time $\tau^{V}_{t-s}$ describes the intervalley exchange scattering of the singlet-triplet trion states. We can analytically solve the system of dynamic equations in the limit of a high doping level and $\tau^{V}_{t-s}=0$. The resulting degree of circular polarization of the spectral lines of the photoluminescence of the exciton is:
\begin{align}
    \eta_{X}&=\frac{\tau^{V}_X \left(n_{e} \tau^r_{X} + C_{b}\right)}{n_{e} \tau^{V}_X \tau^r_{X} + C_{b} \tau^{V}_X + 2 C_{b} \tau^r_{X}}.
\end{align}
Comparing this expression with that obtained for MoSe$_2$ (Eq: \eqref{eq:mose2-high}), we observe an increase in the steady state degree of circular polarization with similar parameters. This difference is due to the presence of two types of trion states in the system of dynamic equations that effectively increase the exciton decay channels. As in the previous case, defining the total exciton decay as $1/\tau_X = 1/\tau^r_X + 2 n_e/C_b$, the solution takes again the form $\eta_{X}=1/(1 + 2 \tau_{X}/\tau^V_X)$. An analytical solution for the low doping level regime can only be obtained for the exciton spectral line, taking again the same functionality. 

\begin{figure}[t!!]
\includegraphics*[keepaspectratio=true, clip=true, angle=0, width=1\columnwidth, trim={3mm, 120mm, 62mm, 0mm}]{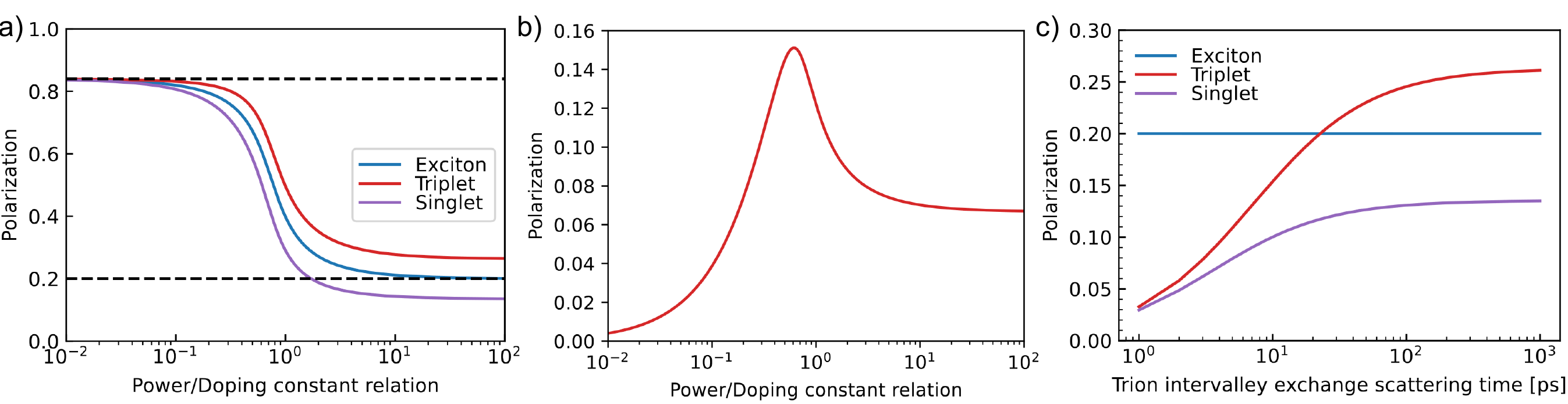}
\caption{\textbf{a)} Degree of circular polarization dependence for triplets (red), singlets (violet) and excitons (blue) as function of the ratio of laser pumping and electron doping level for monolayer WSe$_2$. The horizontal dotted lines correspond to the analytical solutions for weak (upper line) and strong (lower line) free electron level considering $\tau^V_{t-s}=\infty$~ps. \textbf{b)} Degree of free electrons spin-valley polarization as function of the ratio of laser pumping electron doping level in monolayer WSe$_2$ considering $\tau^V_{t-s}=\infty$~ps. \textbf{c)} Degree of circular polarization for triplets, singlets and excitons as function of the trion intervalley exchange scattering time $\tau^{V}_{t-s}$ in monolayer WSe$_2$ considering $P/N_e = 10^2$ (low doping regime). The following parameters were taken as realistic values in \textbf{a}, \textbf{b} and \textbf{c}: $\tau_X=1$~ps, $\tau_{T_t}=2$~ps, $\tau_{T_s}=4$~ps, $\tau^V_X =0.5$~ps and $C_b=0.1\times 10^{10}$~ps\,cm$^{-2}$}
\label{figura:SM-5}
\end{figure}

The numerical solution for WSe$_2$ is presented in figure \ref{figura:SM-5}a. In this case, the circular polarization degree splits in three different curves, one for each excitonic complex. The degree of circular polarization for triplets is higher than the degree of circular polarization for exciton that is also higher than that fir singlets. This splitting is due to different lifetimes of exciton, triplet and singlet states, which induces spin-valley polarization of charge carriers. The dependence of the free charge degree of spin-valley polarization $\eta_{e}=\left({n_{e}^{K}-n_{e}^{K'}}\right)/\left({n_{e}^{K}+n_{e}^{K'}}\right)$ is shown in figure \ref{figura:SM-5}b and display a free electron valley polarization that results from the different exciton and trion valley polarization. $\eta_{e}$ is negligible at low power and high excitation doping level and has a maxima when the excitation power at $K$ is similar to the number of free charges in the system.

Figure \ref{figura:SM-5}c presents the numerical calculation of the degree of circular polarization for WSe$_2$ as function of the intervalley trion scattering $\tau^V_{t-s}$. The circular polarization degree of triplets and singlets display a strong decrease by decreasing $\tau^V_{t-s}$. Thus, the control of the intervalley scattering of trions makes possible to obtain two different regimes when the degree of circular polarization of triplet is higher than that for exciton and vice versa.

\subsection{First principles and time constants calculations}

\subsubsection{Density functional theory calculations}\label{DFT}

We performed density functional theory (DFT) calculations to reveal the microscopic effect of strain on the low energy bands (at the K valleys) that constitute the relevant excitons and trions observed experimentally. We focus on WSe$_2$ because of its larger SOC, thus making it is more susceptible to the effects of strain \cite{zollner2019strain, FariaJunior2022NJP}. The DFT calculations are performed using the WIEN2k code \cite{wien2k}. The crystal structures of strained WSe$_2$ monolayers are generated using the atomic simulation environment (ASE) python package \cite{ASE}. We considered the PBEsol exchange-correlation functional \cite{PerdewPRL2008}, a Monkhorst-Pack k-grid of 15$\times$15 and a self-consistent convergence criteria of 10$^{-2}$ mRy/bohr for the force, 10$^{-6}$ e for the charge, and 10$^{-6}$ Ry for the energy. For the heterostructures. The core–valence energy separation is chosen as $-6$ Ry, the atomic spheres are expanded in orbital quantum numbers up to 10 and the plane-wave cutoff multiplied by the smallest atomic radii is set to 9. For the inclusion of SOC, core electrons are considered fully relativistic whereas valence electrons are treated in a second variational step \cite{Singh2006}, with the scalar-relativistic wave functions calculated in an energy window of -10 to 2 Ry. At zero strain, we found an equilibrium in-plane lattice parameter of 3.2706 $\textrm{\AA}$ and a thickness of 3.3516 $\textrm{\AA}$. We emphasize that for every strain value, the atomic structure is fully relaxed without SOC \cite{zollner2019strain}.

\begin{figure}[t!!]
\includegraphics*[keepaspectratio=true,scale=1.]{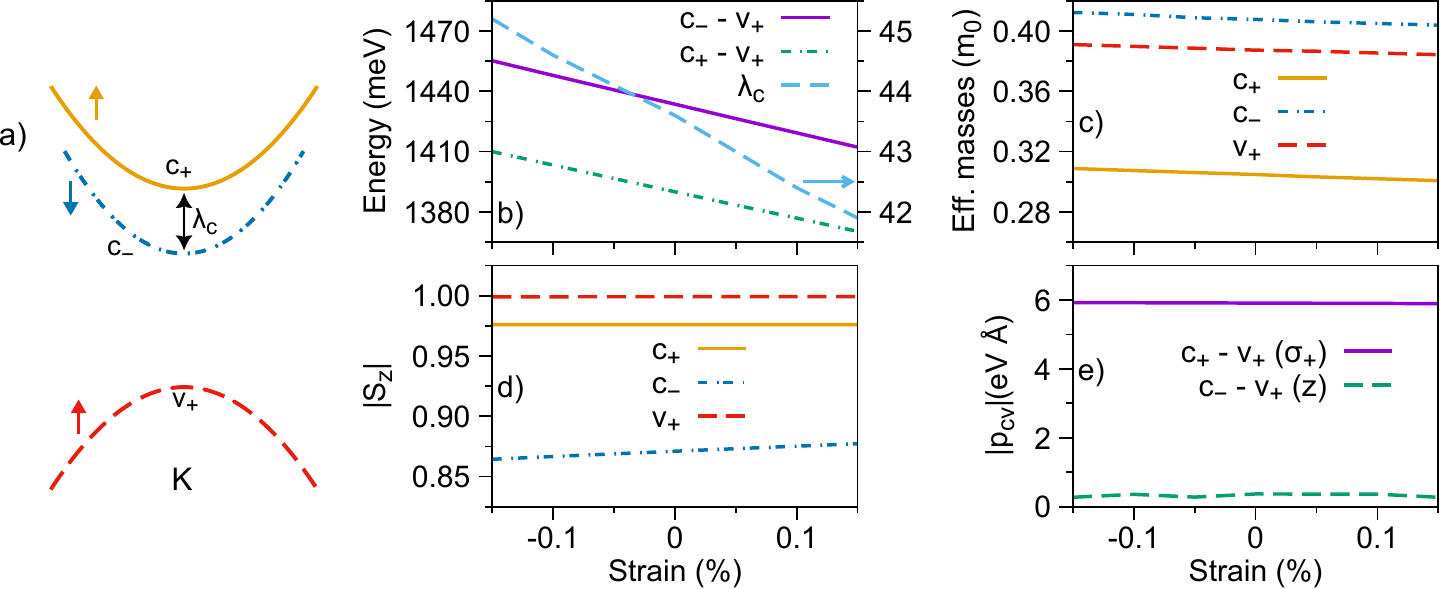}
\caption
{
Summary of DFT results for WSe$_2$ monolayer. \textbf{a)} Schematic representation of the low energy bands relevant to the discussion. The vertical arrows indicate the orientation of the spin [with values given in panel \textbf{d)}]. \textbf{b)} Energy levels, \textbf{c)} effective masses, \textbf{d)} spin expectation value, and \textbf{c)} interband momentum matrix elements as a function of the applied strain. In panel \textbf{b)}, the SOC splitting, $\lambda_c$, is described by the y-axis on the right side.
}
\label{figura:DFT}
\end{figure}

In figure \ref{figura:DFT} we summarize our DFT results. A schematic representation of the low energy bands at the K point is shown in figure \ref{figura:DFT}a. In figure \ref{figura:DFT}b we show the effect of dependence on the transition energies from v$_+$ to c$_\pm$ (left side of y-axis) and the SOC splitting in the conduction band (right side of y-axis). The effective masses for c$_\pm$ and v$_+$ bands as a function of the strain are presented in figure \ref{figura:DFT}c. The spin expectation value, $S_z$, as a function of the strain is shown in figure \ref{figura:DFT}d. Finally, in figure \ref{figura:DFT}e, we present the effect of strain to the interband dipole matrix elements. In summary, our calculations reveal that the transition energies is the most affected quantity, varying on a scale of 30 meV for an applied strain ranging from -0.15\% to +0.15\%. The effective masses, spin expectation value and interband dipole matrix element show a negligible dependence with respect to strain for the analyzed range. Only for large strain values on the order of few \% would introduce non-negligible changes to such these quantities \cite{zollner2019strain, FariaJunior2022NJP}, which are beyond the experimental range in the current manuscript.

\subsubsection{Exciton/trion relaxation times}\label{rad-time}

The total exciton/trion relaxation time ($\tau_{X/T}$) is the reciprocal of the sum of reciprocal decay mechanisms. Considering radiative ($\tau_{X/T}^{rad}$) and non-radiative ($\tau_{X/T}^{non-rad}$) lifetimes, it leads to
\begin{align}
    \tau_{X/T}=\frac{\tau_{X/T}^{rad}\tau_{X/T}^{non-rad}}{\tau_{X/T}^{rad}+\tau_{X/T}^{non-rad}},
\end{align}
Nonradiative scattering includes processes as exciton-phonon, exciton-impurity, and exciton-exciton interaction. In our work, we will not consider in detail the microscopic nature of nonradiative exciton/trion scattering due to the vastness of the problem. However, we calculated the exciton radiative lifetime using the following formula \cite{Glazov2014}:
\begin{align}
    \tau_X^{rad}=\frac{\hbar \varkappa_{b}\omega_X^2m_0^2}{ q e^2 |p_{cv}|^2 |\phi(0)|^2},
\end{align}
where $\varkappa_{b}$ is the environmental high-frequency dielectric constant, $\omega_X$  is the exciton resonance frequency determined by the band gap and binding energy, $p_{cv}$ is the interband momentum matrix element, $\phi(r)$ is the exciton wavefunction and $q=\sqrt{\varkappa_b}\omega_X/c$. We used the value of $\varkappa_{b} = 5$, which corresponds to an hBN encapsulated monolayer. Using ab initio calculations of the band structure of monolayers with a strain \ref{DFT} and variational calculations \cite{Berkelbach2013} to determine the exciton wave function at $r=0$, we calculated the dependence of the exciton radiative time on the strain.
Our results are shown in figure \ref{figura:SM-6}a and display radiative relaxation time that is practically constant under the strain (variation of $\sim 0.5$\,\% along the biaxial strain range).

The trion radiative time can be calculated as follows:
\begin{align}
    \tau^{rad}_T= \frac{\tau^{rad}_X}{4 \pi a_{tr}^2 n_e},
\end{align}
where $a_{tr}$ is the trion radius, which can be estimated using variational approach calculations \cite{Berkelbach2013}. Under the condition of a low doping level (residual doping), we assume that the radiative relaxation time of the trion is much longer than the exciton time \cite{zipfel2020light} (20-40 times longer). The short-range exchange interaction between electrons $V^{exc}_{ee}=\pm U\delta(\vec{r}_1-\vec{r}_2)$ produces a difference in trion radii for the triplet and singlet states \cite{zipfel2020light} which lead to inequality in the radiative times, due to attractive interaction $-U$ for singlet states and repulsive $+U$ for triplet states. Therefore, larger radii lead to faster radiative relaxation time for triplet states than for singlet states.

\subsubsection{Exciton intervalley scattering time}\label{exc-val-time}

\begin{figure}[t!!]
\includegraphics*[keepaspectratio=true, clip=true, angle=0, width=1\columnwidth, trim={1mm, 123mm, 69mm, 0mm}]{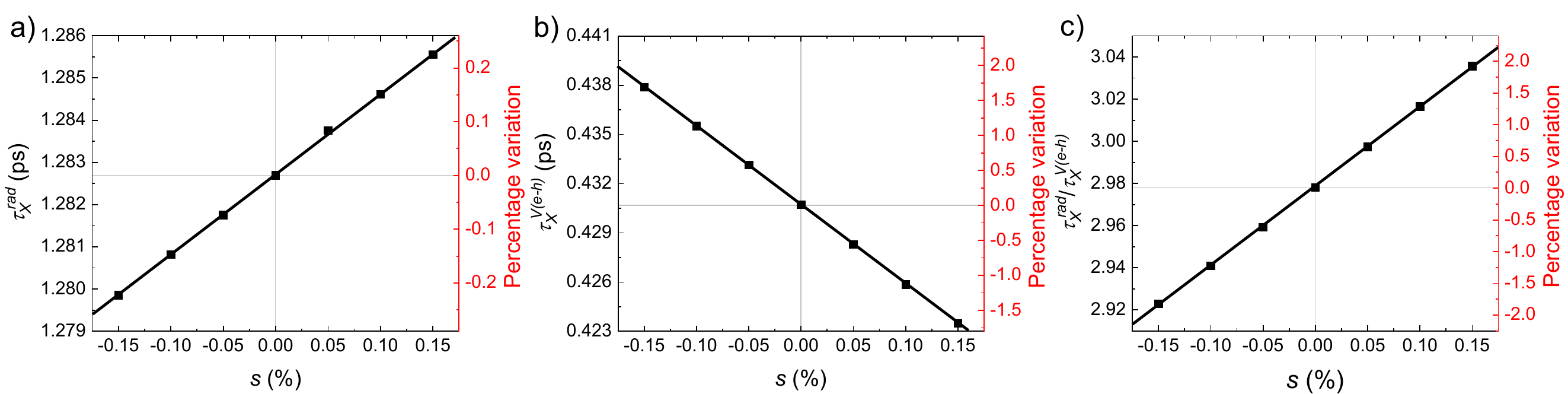}
\caption{\textbf{a)} Calculated radiative exciton relaxation time in monolayer TMDs as function of biaxial strain. The linear dependence is $\tau_X^{rad} = as+b$, where the slope $a=0.015$\,ps/\% is relatively small compared to the offset $b = 1.28$\,ps. \textbf{b)} Calculated exciton valley depolarization time for monolayer TMDs as function of $s$. The linear dependence is $\tau^{V}_X  = as+b$, where the slope $a = -0.05$\,ps/\%. \textbf{c)} Calculated $\tau_X^{rad}/\tau^{V}_X$. The linear dependence is $\tau_X^{rad}/\tau^{V}_X  = as+b$, where the slope $a = 0.377$\,/\%.}
\label{figura:SM-6}
\end{figure}

Exciton intervalley scattering has several sources, such as exciton-exchange \cite{PhysRevB.96.115409}, exciton-phonon \cite{he2020valley}, and exciton-impurity \cite{hao2016direct, cao2012valley} interactions. The exciton-phonon and exciton-impurity interactions require detailed consideration which is beyond the scope of our manuscript. However, considering the exciton-exchange interaction does not require significant calculations. This kind of scattering is present in monolayer TMDs even at zero temperature and in the absence of impurities. The process of valley-to-valley exchange interaction can be considered as a pair electron-hole hopping from one valley to another. We calculate the characteristic time for this process using the following formula \cite{Glazov2014}:
\begin{align}
    \tau^{V(e-h)}_X=\frac{8\hbar\left(q\tau^{rad}_X\right)^2}{M},
\end{align}
where $M=m_e+m_h$ is the exciton effective mass. Note that this expression is quadratic in the parameters $q$ with a direct dependence on $\omega_X$. We calculated the exciton valley depolarization time dependence on the strain, using parameters from ab initio calculations \ref{DFT}. The results are shown in figure \ref{figura:SM-6}b. The exciton exchange valley depolarization time has $\sim 4$\,\% of variation along the biaxial strain range, variation that is an order of magnitude higher than the radiative decay time along same strain range.

Figure \ref{figura:SM-6}c presents the calculated ratio of the radiative decay rate over the EHEI for excitons. This quantity is similar to the ratio of the total decay rate over the total intervalley scattering obtained experimentally and is presented in the main text. Notably, this simple calculation displays a functionality that is similar to the experimental values in MoSe$_2$ and WSe$_2$ monolayers and predicts a strain tuning of the excitonic valley polarization as we demonstrate experimentally. However, the calculation presents for this quotient a slope along the total biaxial strain range that is an order of magnitude smaller than our experimental values.

\subsubsection{Triplet-singlet valley scattering time}\label{trion-val-time}

The triplet to singlet intervalley scattering process is similar to the exciton intervalley scattering process. In the trion case, it consists of a pair of electron-hole hopping from one valley to another, conserving the total momentum and spin. As in the exciton case, several possible channels, such as trion-trion \cite{Kyriienko2020}, trion electron-hole exchange \cite{Courtade2017}, trion-phonon \cite{PhysRevB.106.115407}, and trion-impurity interactions, can contribute to intervalley triplet-singlet scattering of trions. The triplet to singlet scattering time can be estimated using the following formula:
\begin{align}
    \tau^{V(e-h)}_T=\frac{8\hbar\left(q\tau^{rad}_T\right)^2}{M},
\end{align}
where $M=m_e+m_h$ is the exciton effective mass. Considering that the trion radiative relaxation time is longer than the exciton radiative relaxation time, the intervalley scattering of a triplet-singlet pair may be noticeably slower than the intervalley scattering of an exciton. However, the quadratic dependence of the intervalley depolarization time on the radiative time indicates a significant sensitivity of the intervalley scattering of a trion singlet-triplet pair to external control.

\twocolumngrid

%
%

\bibliographystyle{apsrev4-2}
\bibliography{bibliography}

\end{document}